\documentclass[10pt, conference, compsocconf]{IEEEtran}
\ifCLASSINFOpdf
   \usepackage[pdftex]{graphicx}

\else

\fi
\usepackage[cmex10]{amsmath}
\usepackage{hyperref}
\usepackage{xcolor}

\begin{document}
%
\title{Adversarial Attacks in a Multi-view Setting: An Empirical Study of the Adversarial Patches Inter-view Transferability}


 \author{\IEEEauthorblockN{
 Bilel TARCHOUN\IEEEauthorrefmark{1},
 Ihsen ALOUANI\IEEEauthorrefmark{2},   
 Anouar BEN KHALIFA\IEEEauthorrefmark{3},    
 Mohamed Ali MAHJOUB\IEEEauthorrefmark{4}      
 }

  \IEEEauthorblockA{\IEEEauthorrefmark{1}\IEEEauthorrefmark{3}\IEEEauthorrefmark{4}
 Université de Sousse, Ecole Nationale d’Ingénieurs de Sousse, \\ LATIS- Laboratory of Advanced Technology and Intelligent Systems, 4023, Sousse, Tunisie;}
 \IEEEauthorblockA{\IEEEauthorrefmark{2}
 IEMN-DOAE, Université  polytechnique Hauts-de-France, Valenciennes, France}

 Email: \IEEEauthorrefmark{1}tarchoun.bilel@yahoo.com,
  \IEEEauthorrefmark{2}ihsen.alouani@uphf.fr, \\
  \IEEEauthorrefmark{3}anouar.benkhalifa@eniso.u-sousse.tn,
  \IEEEauthorrefmark{4}mohamedali.mahjoub@eniso.rnu.tn

 }


%


\maketitle

\begin{abstract}
While machine learning applications are getting mainstream owing to a demonstrated efficiency in solving complex problems, they suffer from inherent vulnerability to adversarial attacks. Adversarial attacks consist of additive noise to an input which can fool a detector. Recently, successful real-world printable adversarial “patches” were proven efficient against state-of-the-art neural networks. In the transition from digital noise based attacks to real-world physical attacks, the myriad of factors affecting object detection will also affect adversarial patches. Among these factors, view angle is one of the most influential, yet under-explored. In this paper, we study the effect of view angle on the effectiveness of an adversarial patch. To this aim, we propose the first approach that considers a multi-view context by combining existing adversarial patches with a perspective geometric transformation in order to simulate the effect of view angle changes. Our approach has been evaluated on two datasets: the first dataset which contains most real world constraints of a multi-view context, and the second dataset which empirically isolates the effect of view angle. The experiments show that view angle significantly affects the performance of adversarial patches, where in some cases the patch loses most of its effectiveness. We believe that these results motivate taking into account the effect of view angles in future adversarial attacks, and open up new opportunities for adversarial defenses.

\end{abstract}

\begin{IEEEkeywords}
Adversarial Attacks; Perspective Geometric Transformations; View Angles; Adversarial patches; Multi-View; 

\end{IEEEkeywords}

%
\IEEEpeerreviewmaketitle

\begin{figure*}[h]
    \centering
    \includegraphics[scale = 0.69]{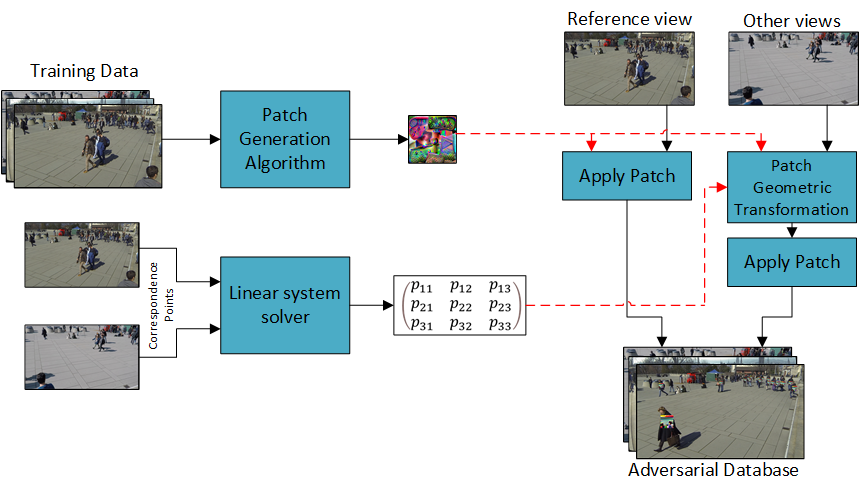}
    \caption{Diagram of the proposed exploration method}
    \label{fig1}
\end{figure*}

\section{Introduction}
Artificial intelligence and machine learning are getting mainstream owing to their efficient problem-solving, and motivated by the need for automatically analyze increasingly larger amounts of data. More specifically, deep neural networks have shown impressive results in image classification and object detection tasks, with many architectures such as Faster R-CNN \cite{1} and YOLO \cite{2} achieving high accuracy and/or real time processing speeds \cite{17}. However, with the increased popularity came increasing scrutiny into the safety and reliability of the results against possible tampering. These suspicions were confirmed by Szegedy \cite{3} and Goodfellow \cite{4} who were able to manipulate the results of deep neural networks with nearly imperceptible adversarial noise. In the hands of a malicious actor, these attacks can be used to fool neural networks, raising security concerns in safety critical applications such as surveillance cameras and self-driving vehicles \cite{16,18}.

\smallskip

Recent work involving adversarial attacks is focusing on applying these attacks in real-world scenarios, and finding defenses against such threats. Since adding noise to an image in real world conditions is impractical, current research is trending towards physical “adversarial patches” \cite{5,6,7}, that can fool detectors and are easy to add to existing objects such as clothing or road signs. Compared to laboratory settings, several factors increase the difficulty of attacking a detector, such as illumination variations, distance from the obstacle and view angles \cite{19}. Moreover, several applications of object detection such as surveillance cameras  \cite{11} and collaborative transportation systems \cite{16} are using multi-view camera systems. With such settings, the current state-of-the-art \cite{8,9} is reaching impressive results even on the most challenging of multi-view datasets \cite{12}.

Although multi-view settings are showing very promising results, very few adversarial attacks consider its effect on their proposed attacks. In fact, patch testing on current state-of-the-art multi-view datasets is still limited to each isolated view. To the best of our knowledge, there is no extensive study in the literature that investigates the impact of multi-view on real-life adversarial attacks. 

In this paper, we study the impact of angle on the effectiveness of adversarial patch attacks. Specifically, we undertake an empirical characterization by generating a patch and applying a perspective geometric transformation to simulate the effects of view angles on the patch. To obtain accurate results, we evaluate our approach on a real-world multi-view dataset, we also introduce a new dataset where the effect of view angle is isolated as much as possible. 

\smallskip

The contributions of our paper can be summarized as follows: \begin{itemize}
    \item We investigate the effect of view-angle on adversarial patches efficiency using geometric transformations. 
    
    \item We introduce a new dataset in which  the factor of angle is isolated.

    \item We empirically evaluate different patches on both our proposed dataset and a multi-view dataset in real-world conditions.
    
    
\end{itemize}

\smallskip

Section 2 of our paper presents our attack approach and setting, Section 3 showcases our experiments and results and discusses the implications of angle on attack effectiveness and we conclude in Section 4.


\section{Proposed methodology}

The aim of our paper is to study the effect of angle on adversarial patches effectiveness. For this goal, we propose an attack setting in which we train an adversarial patch on a multi-view pedestrian detection dataset, then apply the patch to only a single view of the dataset. In the next step, we use perspective geometric transformations to infer the patch on the different views. Finally, we evaluate the patch’s effectiveness by applying a pedestrian detector to the new dataset obtained by applying the patch and subsequent geometric transformations. Figure \ref{fig1} shows a diagram of our method.

In this experiment, we use the attack generation method proposed in \cite{5}, as it has shown good performance especially when applied to a objects with high intra-class variations such as persons. The goal of this attack is to generate a patch that hides people from the detector by minimizing the objectness score of the person to hide in the context of surveillance camera systems. The patch is created by optimizing the pixels in the patch area using an Adam optimizer that minimizes the following loss function shown in equation \ref{eq1}:

\begin{equation}
    L= \alpha L_{nps} + \beta L_{tv} + \gamma L_{obj}
    \label{eq1}
\end{equation}

\smallskip

Where $\alpha$, $\beta$ and $\gamma$ are empirically determined scaling factors, $L_{nps}$ is a non-printability score that ensures that the colors in the patch are printable by common printers, $L_{tv}$ symbolizes the difference in colors between neighboring pixels, and emphasizes smoother patches, and $L_{obj}$ is the maximum objectness score in the image. This is the main component in the loss function, as the lower this score is, the less likely the detector is able to detect the person.

After generating the patch, we apply it to a single view of the dataset (henceforth referred to as the “reference view"). Then, we apply a perspective geometric transformation (P) between each frame of the reference view and its corresponding frame in other views using the geometric transformation described in Equation \ref{eq2}: 

\smallskip

\begin{equation}
    \begin{pmatrix}
        x^{dst} \\
        y^{dst} \\
        1
    \end{pmatrix}
    =
    \begin{pmatrix}
        p_{11} & p_{12} & p_{13}\\
        p_{21} & p_{22} & p_{23} \\
        p_{31} & p_{32} & 1
    \end{pmatrix}
    \begin{pmatrix}
        x^{ref} \\
        y^{ref} \\
        1
    \end{pmatrix}
    \label{eq2}
\end{equation}

Where $(x^{ref},y^{ref})$ refers to a pixel in the reference view,  $(x^{dst},y^{dst})$ refers to a pixel in the destination view, $(p_{11}, p_{12}, p_{21}, p_{22})$ form the rotation matrix and are composed of various transformations to the image (Scale, shear and rotation), $p_{31}$ and $p_{32}$ are the translation vector, and $p_{13}$ and $p_{23}$ form the tilt vector. Obtaining the parameters of the geometric transformation can be likened to solving the linear system shown in equation \ref{eq3}, requiring at least four correspondence points that can be obtained using calibration data or manually extracted. 

\smallskip

\begin{equation}
    \left\{
    \begin{matrix}
        \displaystyle x^{dst} = \frac{p_{11}*{x_{ref}}+p_{12}*{y_{ref}}+p_{13}}{p_{31}*x_{ref}+p_{32}*{y_{ref}}} \\
        \displaystyle x^{ref} = \frac{p_{21}*{x_{ref}}+p_{22}*{y_{ref}}+p_{23}}{p_{31}*{x_{ref}}+p_{32}*{y_{ref}}}
    \end{matrix}
    \right.
    \label{eq3}
\end{equation}

Finally, we obtain a new dataset with an adversarial patch applied to the reference view and transferred to other views via perspective geometric transformations. We can now apply the chosen pedestrian detector to the dataset we have prepared.

\section{Experiments and results}
In this section we apply then aforementioned method in the previous section and discuss the obtained results:

\subsection{Testing Parameters}

In order to generate the patch, we use the following empirical hyperparameters shown in Table \ref{tab1}. All trained patches use random values as a starting point.

\smallskip

\begin{table}[h]
    \centering
    \caption{Patch training hyperparameters}
    \begin{tabular}{|l|c|}
        \hline
            Hyperparameter & Value  \\
        \hline
            $\alpha$ & 0.01 \\
        \hline
            $\beta$ & 2.5* \\
        \hline
            $\gamma$ & 1 \\
        \hline
            Patch size & 300*300 px \\
        \hline
            Mini batch size & 4 \\
        \hline
            Learning rate & 0.03 \\
        \hline
        \multicolumn{2}{c}{*$L_{tv}$ is floored at 0.1 to prevent the optimizer from focusing} \\
        \multicolumn{2}{c}{on smoothing the patch over achieving high fooling rates} 
    \end{tabular}
    \label{tab1}
\end{table}

To ensure proper testing conditions, we choose YOLOv2 for pedestrian detection, as it was the target detector of the chosen adversarial patch. Furthermore, we use pre-trained weights for the YOLOv2 detector that are provided by the authors of the patch.

\subsection{Wildtrack tests and results}
\begin{figure}[tp]
    \centering
    \includegraphics[scale = 0.25]{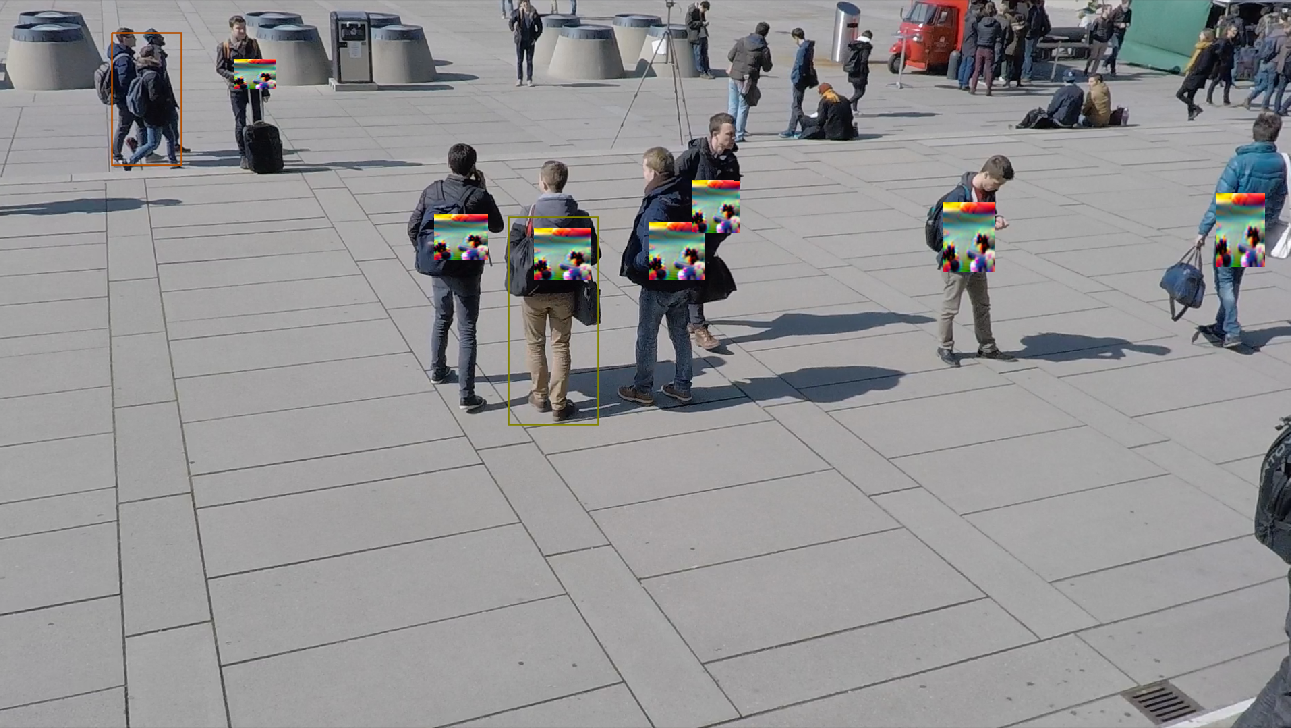}
    \caption{Sample of the Wildtrack view 7 detection results with the patch applied after geometric transformation}
    \label{fig4}
\end{figure}

We first test our patch on the Wildtrack multiview dataset \cite{11}. We generate our patch on view $1$, and we divide the views into two sets $\{1,4,6,7\}$ and $\{2,3,5\}$ since we add a patch to each visible person’s front and back. We choose views $1$ and $5$ as reference views since those cameras are opposed to each other. Figure \ref{fig4} shows the patch added in a sample frame of view $7$ after the geometric transformation step. The results of these experiments are shown in tables \ref{tab2} and \ref{tab3} below. The recall values are the percentage of people correctly detected among the number of people that are present in both the reference view and destination view since we do not take into account people who are only visible in only one of the views.


\begin{table}[h!]
    \centering
    \caption{Patch performance on the \{1,4,6,7\} view set}
    \begin{tabular}{|l|l|l|l|}
         \hline
         View & Clean Recall & Patched Recall & Difference (\%)  \\
         \hline
         1 (ref) & 17.35 & 1.04 & -94.01\% \\ 
         \hline
         4 & 30.28 & 15.86 & -47.62\% \\ 
         \hline
         6 & 7.56 & 4.03 & -46.69\% \\ 
         \hline
         7 & 44.22 & 6.97 & -84.24\% \\ 
         \hline
    \end{tabular}
    \label{tab2}
\end{table}

\begin{table}[h!]
    \centering
    \caption{Patch performance on the \{2,3,5\} view set}
    \begin{tabular}{|l|l|l|l|}
         \hline
         View & Clean Recall & Patched Recall & Difference (\%)  \\
         \hline
         5 (ref) & 43.37 & 9.72 & -77.59\% \\ 
         \hline
         2 & 22.97 & 20.85 & -9.23\% \\ 
         \hline
         3 & 24.59 & 10.28 & -58.19\% \\ 
         \hline
    \end{tabular}
    \label{tab3}
\end{table}

\newpage

\subsection{LATIS-MVAPE database tests and results}

To better isolate the effects of view angle, we create our own dataset: LATIS- Multi-View Adversarial Patch Evaluation (LATIS-MVAPE) which contains $14$ images of $18$ different persons taken from seven different view angles and two distances for a total of $252$ photos. $11$ persons were filmed in indoor conditions, and the other seven were filmed outdoors. For the geometric transformation correspondence points, we manually select $18$ pairs of points between the reference view and each destination view. We train 2 patches: The first patch with only the reference view as input data, and the second patch with the reference and  $60 ^{\circ}$  views from the dataset as input data. Figure \ref{fig6} shows a sample of patch application with the subsequent geometric transformation in both the reference view and in the $60 ^{\circ}$ view respectively. Table \ref{tab4} below shows the detection results with each of the applied patches:

\smallskip

\begin{figure}[tp]
    \centering
    \includegraphics[scale = 0.50]{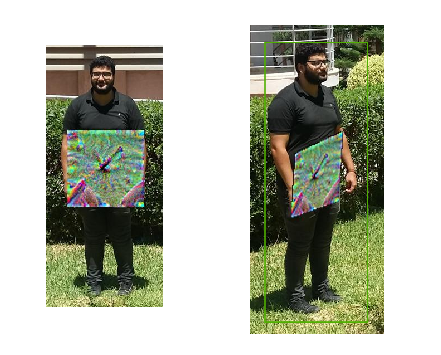}
    \caption{Sample of a successful adversarial attack on the reference view of LATIS-MVAPE, while it fails to fool the detector at a different angle.}
    \label{fig6}
\end{figure}



\smallskip

\begin{table}[h!]
    \centering
    \caption{Detection results using the two patches trained on the LATIS-MVAPE dataset}
    \begin{tabular}{|l|l|l|l|}
         \hline
         View & Clean Recall & Patch 1 Recall & Patch 2 Recall  \\
         \hline
         $5^{\circ}$ & 100 & 70.59 & 64.71 \\ 
         \hline
         $30^{\circ}$ & 100 & 75.00 & 72.22 \\ 
         \hline
         $60^{\circ}$ & 100 & 91.67 & 100.00  \\ 
         \hline
        $ 90^{\circ}$ (ref) & 100 & 38.89 & 27.78 \\ 
         \hline
        $ 120^{\circ}$ & 100 & 92.86 & 96.43  \\ 
         \hline
         $150^{\circ}$ & 100 & 92.86 & 82.14  \\ 
         \hline
         $175^{\circ}$ & 100 & 83.33 & 83.33  \\ 
         \hline
    \end{tabular}
    \label{tab4}
\end{table}

\subsection{Discussion}

Our results confirm that angle has a significant effect on the effectiveness of an adversarial patch attack: In Wildtrack database experiments, the projected patch lost up to half its efficiency compared to the one applied to the reference view. In the tests performed on our database, the projected patch was at best half as effective as the reference view patch, losing nearly all of its attack capabilities in some cases. 


These findings have the potential to impact both the attacking side and the defending side in adversarial patch attacks: On the attack side, this study highlights the need to incorporate view angle into the patch creation process, as the patch loses a significant amount of its effectiveness when viewed from an angle. This integration can be achieved by adding geometric
transformations in the patch training process, or including multi-view data in the training dataset.

As for the defense side, these findings suggest interesting pathways to stronger defenses, such as adopting multi-view detection techniques. These multi-view detectors combine data from all of the available views to locate objects within them, and as such can detect objects in a certain view that a monocular detector failed to detect using the extra multi-view data. Therefore, to attack such a potential defense, an adversary has to defeat the detector on all of the views, reducing the strength of the adversarial patch to be only as strong as the weakest view, and in the case of a patch untrained against view angle variations, this effectiveness loss can be severe.

\section{Conclusion}
The threat of adversarial attacks has spurred increasing interest in designing practical adversarial attacks and defenses in the real world. To better understand these attacks, an investigation of the interactions between adversarial patches and the factors that affect object detection is necessary. This paper unprecedentedly studies the effect of view angle on the effectiveness of an adversarial patch. To emulate the angle view deformation on the patch, we apply a perspective geometric transformation to an existing attack. We compared the original patch results with the transformed patch on other views, and we notice a significant effect on attack success rates. This difference has implications for adversarial patches on both tasks of attacking a detector or immunizing detectors from such attacks. 





%

\def\IEEEbibitemsep{0pt}

\bibliographystyle{IEEEtran}
\bibliography{ref.bib}

\end{document}